\documentclass
[preprintnumbers,floats,noshowpacs,bibnotes,10pt,twocolumn,prl]{revtex4}%
\usepackage{graphicx}
\usepackage{amsmath}
\usepackage{amsfonts}
\usepackage{amssymb}%
\providecommand{\U}[1]{\protect\rule{.1in}{.1in}}
\providecommand{\U}[1]{\protect\rule{.1in}{.1in}}
\providecommand{\U}[1]{\protect\rule{.1in}{.1in}}
\begin{document}
\preprint{ }
\title{Stochastic Resonance with a Single Metastable State}
\author{Eran Segev}
\email{segeve@tx.technion.ac.il}
\author{Baleegh Abdo}
\author{Oleg Shtempluck}
\author{Eyal Buks}
\affiliation{Department of Electrical Engineering, Technion, Haifa 32000, Israel}
\date{\today}

\begin{abstract}
We study thermal instability in NbN superconducting stripline resonators. The
system exhibits extreme nonlinearity near a bifurcation, which separates a
monostable zone and an astable one. The lifetime of the metastable state,
which is locally stable in the monostable zone, is measure near the
bifurcation and the results are compared with a theory. Near bifurcation,
where the lifetime becomes relatively short, the system exhibits strong
amplification of a weak input modulation signal. We find that the frequency
bandwidth of this amplification mechanism is limited by the rate of thermal
relaxation. When the frequency of the input modulation signal becomes
comparable or larger than this rate the response of the system exhibits
sub-harmonics of various orders.

\end{abstract}

\maketitle

Stochastic resonance (SR) is a phenomenon in which metastability in nonlinear
systems is exploited to achieve amplification of weak signals
\cite{stocRes_Benzi83,stocRes_Gammaitoni98,stocRes_Wellens04}. SR has been
experimentally demonstrated in electrical, optical, superconducting, neuronal
and mechanical systems
\cite{stocRes_Fauve83,stocRes_McNamara88,stocRes_Rouse95,stocRes_Hibbs95,Baleegh06c,stocRes_Longtin91,stocRes_Levin96,stocRes_Badzey05,stocRes_Chan06}%
. Usually, SR is achieved by operating the system in a region in which it has
more than one locally stable (metastable) steady state and its response
exhibits hysteresis. Under some appropriate conditions, a weak input signal,
which modulates the transition rates between these states, can lead to
synchronized noise-induced transitions, allowing thus strong amplification.

In the present paper we investigate SR and amplification in a superconducting
(SC) NbN stripline resonator. Contrary to previous studies, we operate the
system near a bifurcation between a monostable zone, in which the system has a
single metastable state, and an astable zone, in which this state ceases to
exist and the system lacks any steady states. In our previous studies we have
investigated several effects, e.g. strong amplification \cite{segev06d}, noise
squeezing \cite{segev06d}, and response to optical illumination
\cite{Segev06a,segev06e}, which occur near this bifurcation, and limit cycle
oscillations, which are observed in the astable zone \cite{segev06b,segev06c}.
In the present work we investigate experimentally and theoretically the
response of the system to amplitude modulated input signal, and find an
unusual SR mechanism that has both properties of strong responsivity and
non-hysteretic behavior. The frequency bandwidth of this mechanism is found to
be limited by the rate of thermal relaxation. We find that rather unique
sub-harmonics of various orders are generated when the modulation frequency
becomes comparable or larger than the relaxation rate. Moreover, we measure
the lifetime of the metastable state in the monostable zone near the
bifurcation and compare the results with a theory.

\begin{figure}
[ptb]
\begin{center}
\includegraphics[
height=1.1083in,
width=3.4006in
]%
{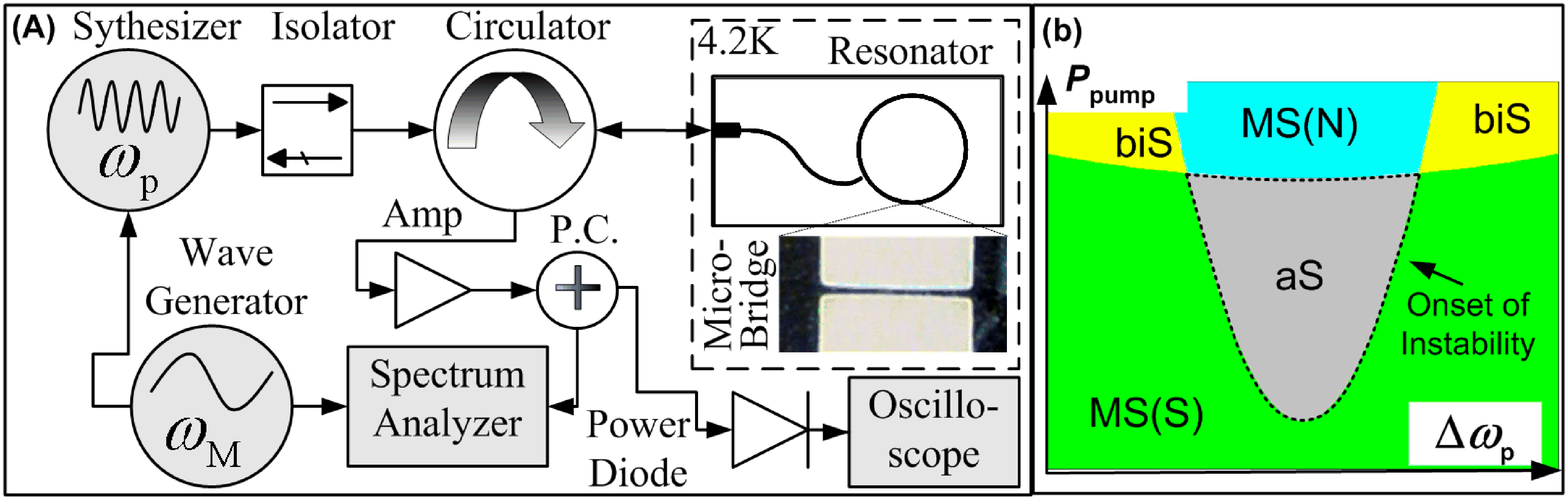}%
\caption{$\left(  \mathrm{A}\right)  $ Experimental setup. $\left(
\mathrm{B}\right)  $ System stability diagram.}%
\label{expSetupAndSM}%
\end{center}
\end{figure}

Our experiments are performed using a novel device that integrates a narrow
microbridge into a SC stripline electromagnetic resonator (see Fig.
\ref{expSetupAndSM} $\left(  \mathrm{A}\right)  $). Design considerations,
fabrication details as well as resonance modes calculation can be found
elsewhere \cite{Segev06a}. The dynamics of our system can be captured by two
coupled equations of motion, which are hereby briefly described (see Ref.
\cite{segev06c} for a detailed derivation). Consider a resonator driven by a
weakly coupled feed-line carrying an incident amplitude modulated coherent
tone $b^{\mathrm{in}}=b_{0}^{\mathrm{in}}(1+a\cos(\omega_{\mathrm{m}%
}t))e^{-i\omega_{\mathrm{p}}t}$, where $b_{0}^{\mathrm{in}}$ is constant
complex amplitude, $\omega_{\mathrm{p}}$ is the driving angular frequency, $a$
is the modulation depth, and $\omega_{\mathrm{m}}\ll$ $\omega_{\mathrm{p}}$ is
the modulation frequency.\ The mode amplitude inside the resonator can be
written as $Be^{-i\omega_{\mathrm{p}}t}$, where $B\left(  t\right)  $ is a
complex amplitude, which is assumed to vary slowly on a time scale of
$1/\omega_{\mathrm{p}}$. \ In this approximation, the equation of motion of
$B$ reads \cite{Squeezing_Yurke05}%
\begin{equation}
\frac{\mathrm{d}B}{\mathrm{d}t}=\left[  i\left(  \omega_{\mathrm{p}}%
-\omega_{0}\right)  -\gamma\right]  B-i\sqrt{2\gamma_{1}}b^{\mathrm{in}%
}+c^{\mathrm{in}},\label{dB/dt}%
\end{equation}
where $\omega_{0}$ is the angular resonance frequency and $\gamma\left(
T\right)  =\gamma_{1}+\gamma_{2}\left(  T\right)  $, where $\gamma_{1}$ is the
coupling coefficient between the resonator and the feed-line and $\gamma
_{2}\left(  T\right)  $ is the temperature dependant damping rate of the mode,
and $T$ is the temperature of the microbridge.\ The term $c^{\mathrm{in}}$
represents an input Gaussian noise. The microbridge heat balance equation reads%

\begin{equation}
C\frac{\mathrm{d}T}{\mathrm{d}t}=2\hslash\omega_{0}\gamma_{2}\left\vert
B\right\vert ^{2}-H\left(  T-T_{0}\right)  , \label{dT/dt}%
\end{equation}
where $C$ is the thermal heat capacity, $H$ is the heat transfer coefficient,
and $T_{0}=4.2%
\operatorname{K}%
$ is the temperature of the coolant.

Coupling between Eqs. (\ref{dB/dt}) and (\ref{dT/dt}) originates by the
dependence of the damping rate $\gamma_{2}\left(  T\right)  $ of the driven
mode on the resistance of the microbridge \cite{supRes_Saeedkia05}, which in
turn depends on its temperature. We assume the simplest case, where this
dependence is a step function that occurs at the critical temperature
$T_{\mathrm{c}}\simeq10%
\operatorname{K}%
$ of the superconductor, namely $\gamma_{2}$ takes the value $\gamma
_{2\mathrm{s}}$ for the SC $T<T_{\mathrm{c}}$ phase of the microbridge and
$\gamma_{2\mathrm{n}}$ for the normal-conducting (NC) $T>T_{\mathrm{c}}$
phase.%

Solutions of steady state response to a monochromatic excitation (no
modulation $a=0$) are found by seeking stationary solutions to Eqs.
(\ref{dB/dt}) and (\ref{dT/dt}) for the noiseless case $c^{\mathrm{in}}=0$.
Due to the coupling the system may have, in general, up to two locally-stable
steady-states, corresponding to the SC and NC phases of the microbridge. The
stability of each of these phases depends on both the power, $P_{\mathrm{pump}%
}\propto$ $\left\vert b^{\mathrm{in}}\right\vert ^{2}$, and frequency
$\omega_{p}$ parameters of the injected pump tone. Our system has four
stability zones (Fig. \ref{expSetupAndSM}$\left(  \mathrm{b}\right)  $)
\cite{segev06c}. Two are mono-stable zones (MS(S) and MS(N)), where either the
SC or the NC phases is locally stable, respectively. Another is a bistable
zone (BiS), where both phases are locally stable
\cite{Baleegh_bifurcation,Baleegh06a}. The third is an astable zone (aS),
where none of the phases are locally stable. Consequently, when the resonator
is biased to this zone, the microbridge oscillates between the two phases. The
onset of this instability, namely the bifurcation threshold (BT), is defined
as the boundary of the astable zone (see Fig. \ref{expSetupAndSM}$\left(
\mathrm{b}\right)  $).

The experimental setup is depicted in Fig. \ref{expSetupAndSM}$(\mathrm{a})$.
We inject an amplitude modulated pump tone into the resonator and measure the
reflected power in the frequency domain using a spectrum analyzer and in the
time domain using an oscilloscope. The parameters used for the numerical
simulation were obtained as follows. The coupling coefficient $\gamma
_{1}=2\operatorname{MHz}$ and the damping rates $\gamma_{2\mathrm{s}}=$
$2.2\operatorname{MHz}$, $\gamma_{2\mathrm{n}}=64\operatorname{MHz}$ were
extracted from frequency response measurement
\cite{Segev06a,Baleegh_bifurcation}, whereas the thermal heat capacity
$C=54\operatorname{nJ}\operatorname{cm}^{-2}\operatorname{K}^{-1}$ and the
heat transfer coefficient\ $H=12\operatorname{W}\operatorname{cm}%
^{-2}\operatorname{K}^{-1}$ were calculated analytically according to Refs.
\cite{kinInd_Johnson96,HED_Weiser81}.
\begin{figure}
\centering
$%
\begin{array}
[c]{c}%
\text{%
{\parbox[b]{3.3723in}{\begin{center}
\includegraphics[
height=1.2611in,
width=3.3723in
]%
{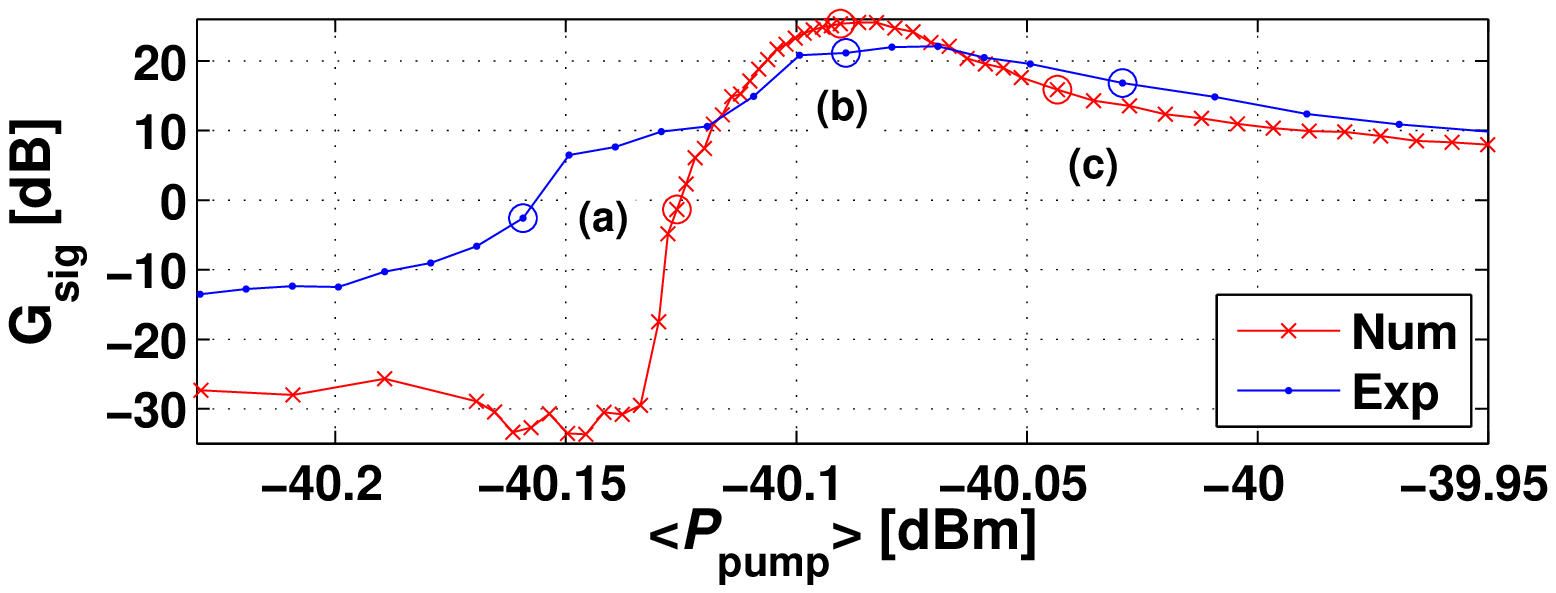}%
\\
$\left(  \mathrm{A}\right)  $%
\end{center}}}%
}\\
\text{%
{\parbox[b]{3.3723in}{\begin{center}
\includegraphics[
height=2.5313in,
width=3.3723in
]%
{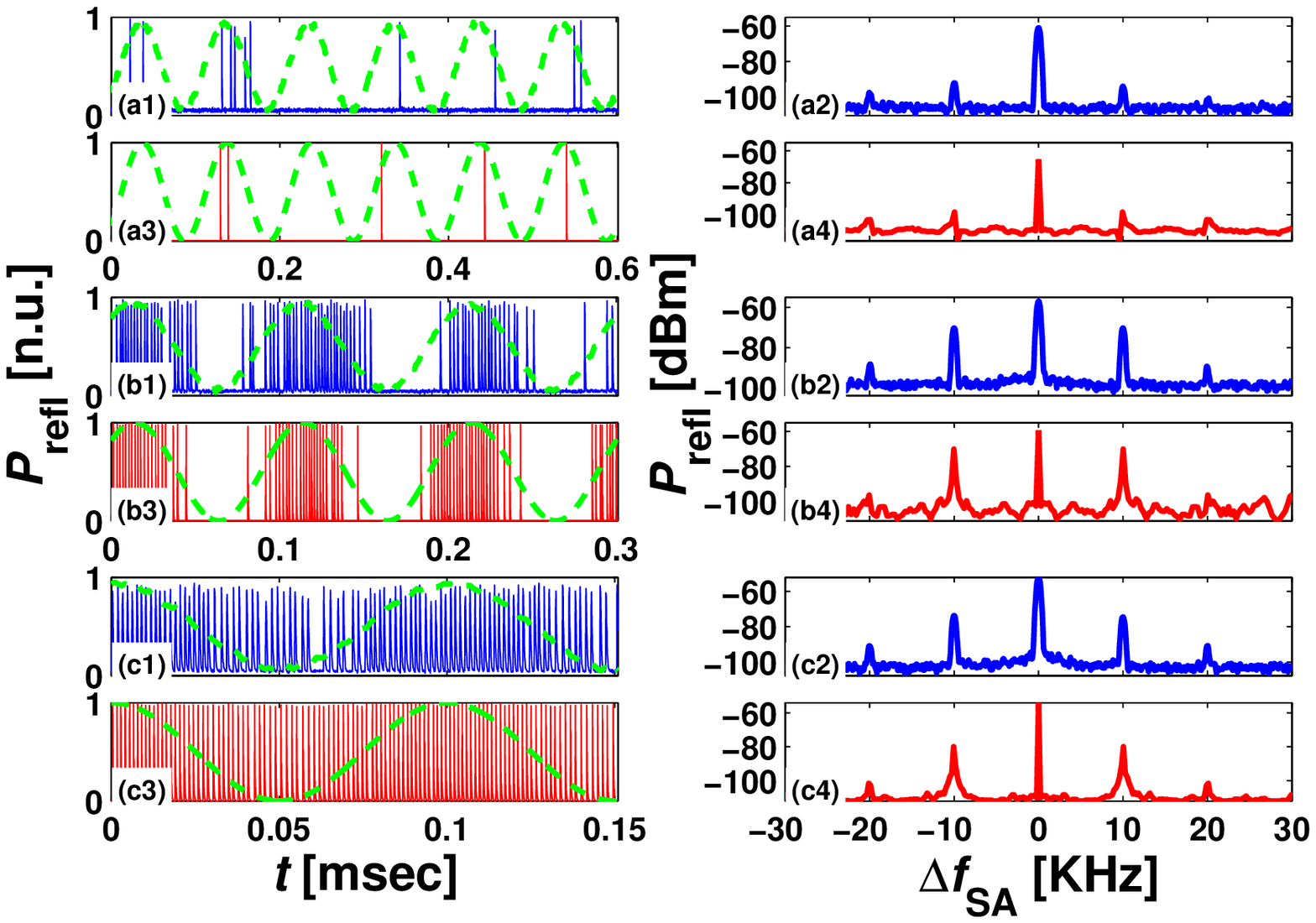}%
\\
$\left(  \mathrm{B}\right)  $%
\end{center}}}%
}%
\end{array}
$%
\caption{$\left(  \mathrm{A}\right)  $ Experimental (dotted-blue) and
numerical (crossed-red) results of the signal amplification \textrm
{G}$_{\mathrm{sig}}$ as a function of the mean injected pump power
$<P_{\mathrm{pump}}>$.
$\left(  \mathrm{B}\right)  $ Experimental (subplots $\left(  x{\small1}%
\right)  $ and $\left(  x{\small2}\right)  $) and numerical (subplots
$\left(  x{\small3}\right)  $ and $\left(  x{\small4}\right)  $) results of
the reflected power $P_{\mathrm{refl}}$ as a function of time (subplots
$\left(  x{\small1}\right)  $ and $\left(  x{\small3}\right)  $) and scanned
frequency $f_{\mathrm{SA}}$ (subplots $\left(  x{\small2}\right)  $ and
$\left(  x{\small4}\right)  $), centered on the resonance frequency
$f_{0}=4.363\mathrm{GHz}$ ($\Delta f_{\mathrm{SA}}=f_{\mathrm{SA}}%
-f_{0}$), where $x$ denotes $a$, $b$, and $c$,
corresponding to the marked points in panel $\left(  \mathrm{A}\right) $. The
dashed-green curve represents the modulation signal. The time domain
measurements are normalized by their maximum peak to peak value.}%
\label{AmplitudeMod10KHz}%
\end{figure}%

Our system exhibits an extremely strong amplification when tuned to the BT.
Figure \ref{AmplitudeMod10KHz} shows both experimental (Blue curves) and
numerical (Red curves) results for the case where the system is driven by a
modulated pump tone having the following parameters: $\omega_{\mathrm{p}%
}=\omega_{0}=2\pi\times4.363%
\operatorname{GHz}%
$, $\omega_{\mathrm{m}}=2\pi\times10%
\operatorname{kHz}%
$, $a=0.0024$, and by an effective noise temperature of $T_{\mathrm{eff}}=75%
\operatorname{K}%
$. Panel $\left(  \mathrm{A}\right)  $ plots the signal gain $G_{\mathrm{sig}%
}$, defined as the ratio between the reflected power at frequency
$\omega_{\mathrm{p}}+\omega_{\mathrm{m}}$ and the sum of the injected powers
at frequencies $\omega_{\mathrm{p}}\pm\omega_{\mathrm{m}}$, as a function of
the mean injected pump power $\left\langle P_{\mathrm{pump}}\right\rangle $.
The system exhibits large gain of approximately $20\mathrm{dB}$ around the BT.
The experimental results exhibits excess gain below BT relative to the
numerical results. This can be explained by additional nonlinear mechanisms
\cite{supRes_Golosovsky95} that may induce small amplification, and are not
theoretically included in our piecewise linear model.

Figure \ref{AmplitudeMod10KHz} $\left(  \mathrm{B}\right)  $, shows time and
frequency domain results of the reflected power, for three pairs of input
power values, corresponding to the marked points $\left(  \mathrm{a-c}\right)
$ in panel $\left(  \mathrm{A}\right)  $. In addition, the time domain
measurements contain a green-dashed curve showing the modulating signal. The
results shown in subplots $\left(  \mathrm{a1-a4}\right)  $ were obtained
while biasing the system below the BT, namely, $\left\langle P_{\mathrm{pump}%
}\right\rangle $ was set below the power threshold, $P_{\mathrm{c}}$. In
general, the spikes in the time domain plots of Fig. \ref{AmplitudeMod10KHz}
$\left(  \mathrm{B}\right)  $ indicate events in which the temperature $T$
temporarily exceeds $T_{\mathrm{c}}$ \cite{segev06c}. Below threshold, the
average time between such events, which are induced by input noise, is the
lifetime $\Gamma^{-1}$ of the metastable state of the resonator. As we will
show in the last part of this paper, $\Gamma$ strongly depends on the pump
power near BT, thus power modulation results in a modulation of the rate of
spikes, as can be seen both in the experimental and simulation results.

Subplots $\left(  \mathrm{b1-b4}\right)  $ of Fig. \ref{AmplitudeMod10KHz}%
$\left(  \mathrm{B}\right)  $ show experiments in which $\left\langle
P_{\mathrm{pump}}\right\rangle \simeq P_{\mathrm{c}}$ and thus, the modulation
itself drives the resonator in and out the astable zone. As a result, during
approximately half of the modulation period nearly regular spikes in reflected
power are observed, whereas during the other half only few noise-induced
spikes are triggered. This behavior leads to a very strong gain as well as to
the creation of higher order frequency components (subplots $\left(
\mathrm{b2,b4}\right)  $). Figure \ref{AmplitudeMod10KHz}$\left(
\mathrm{B}\right)  $, Subplots $\left(  \mathrm{c1-c4}\right)  $, show
experiments in which $\left\langle P_{\mathrm{pump}}\right\rangle
>P_{\mathrm{c}}$, and thus the regular spikes occur throughout the modulation
period. The rate of the spikes is strongly correlated to the injected power
\cite{segev06b}, and it is higher for stronger pump powers. \ Therefore, as
the injected pump power is modulated, so is that rate. This behavior also
creates a rather strong amplification, though weaker than the one achieved in
the previous case.

The amplification mechanism in our system is unique in several aspects. First
it is extremely strong. To emphasize the strength of the amplification we note
that, usually, no amplification greater than unity ($0~$dB) is achieved in
such measurements with SC resonators \cite{Nonlinear_Chin92,IM_Monaco2000},
unless the resonator is driven near BT \cite{BifAmp_Erik07}. In addition, it
does not exhibit a hysteretic behavior.
\begin{figure}
\centering
$%
\begin{array}
[c]{c}%
\text{%
{\parbox[b]{3.4014in}{\begin{center}
\includegraphics[
height=2.5604in,
width=3.4014in
]%
{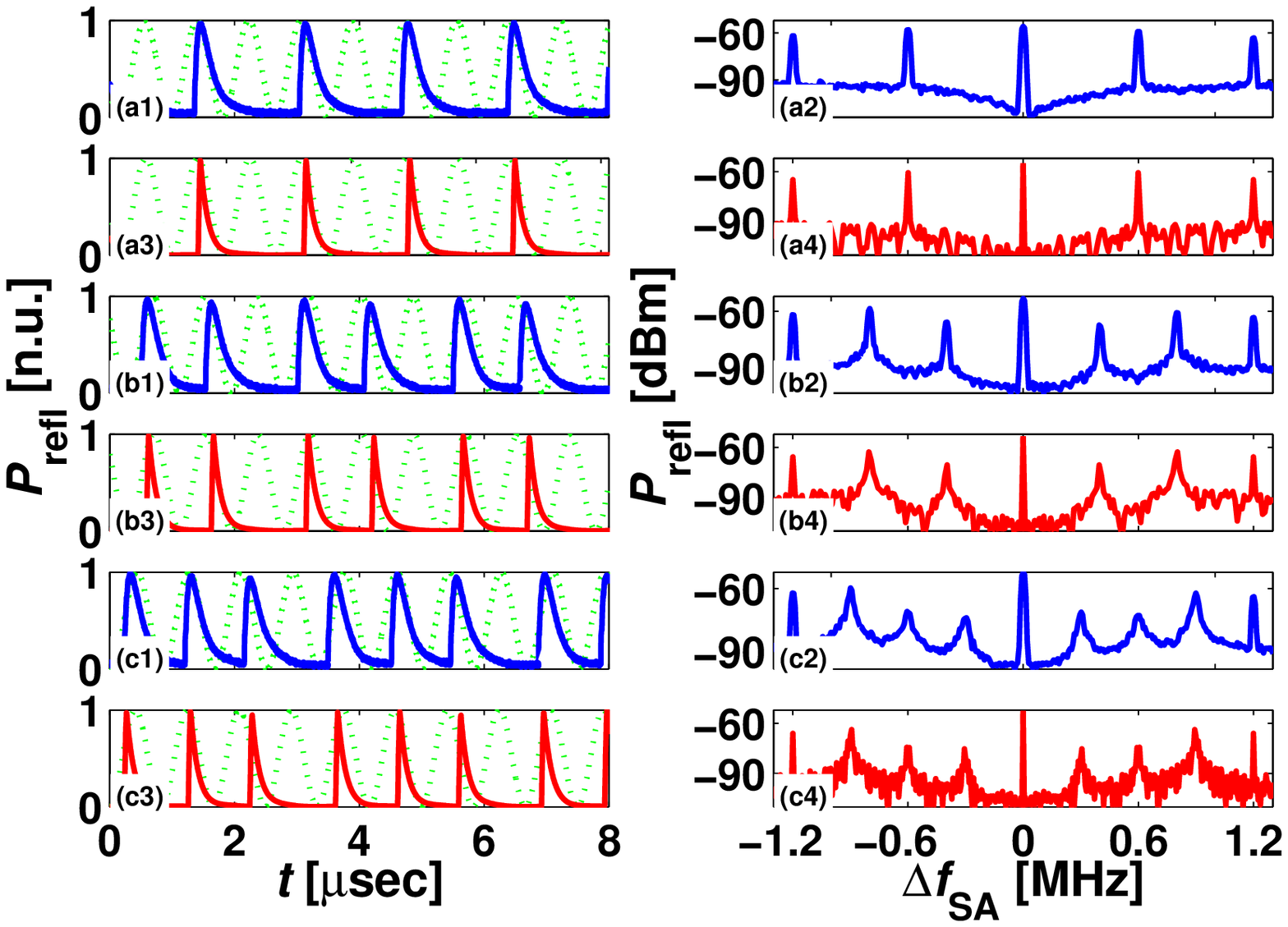}%
\\
$\left(  \mathrm{A}\right)  $%
\end{center}}}%
}\\
\text{%
{\parbox[b]{3.3723in}{\begin{center}
\includegraphics[
height=1.0502in,
width=3.3723in
]%
{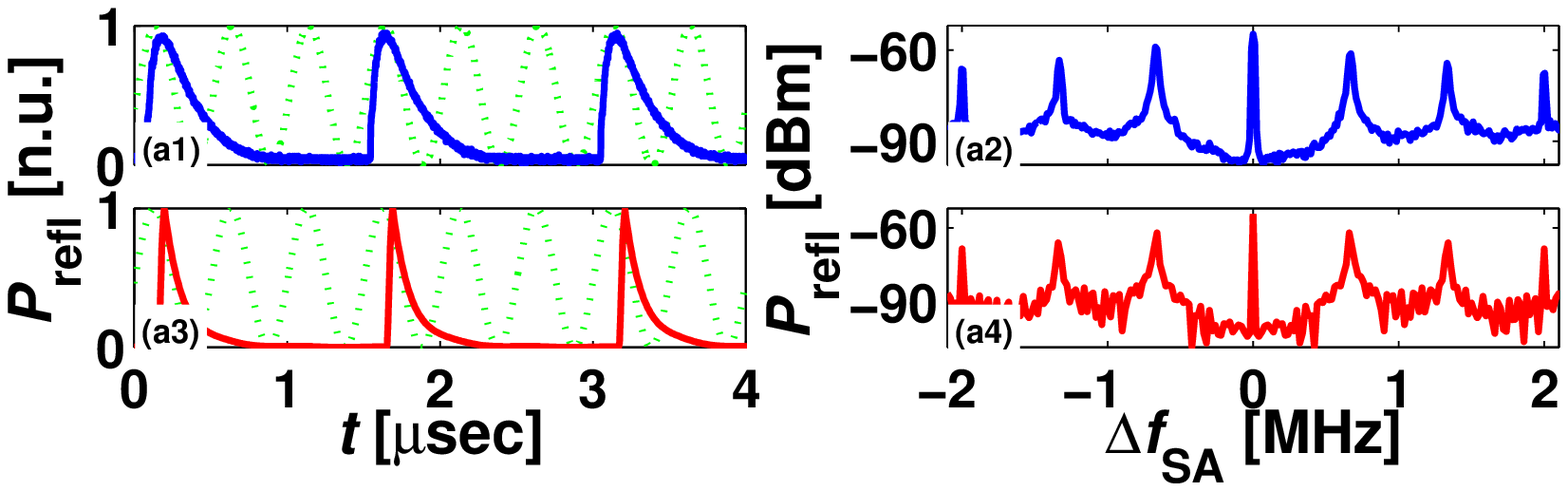}%
\\
$\left(  \mathrm{B}\right)  $%
\end{center}}}%
}%
\end{array}
$%
\caption{Sub-harmonics generation.}%
\label{SubHarmLocking}%
\end{figure}%

Each spike in subplots $\left(  \mathrm{a1-a2}\right)  $ of Fig.
\ref{AmplitudeMod10KHz}$\left(  \mathrm{B}\right)  $ lasts approximately $1%
\operatorname{\mu s}%
$, after which the device is ready to detect a new event. This recovery time
determines the detection bandwidth. A measurement of the dependence of the
amplification mechanism on the modulation frequency $\omega_{\mathrm{m}}$ has
reveled a mechanism in which sub-harmonics of the modulation frequency are
generated by the device. The generation occurs when the modulation period is
comparable to the recovery time of the system. The results are shown in Fig.
\ref{SubHarmLocking} which shows both experimental (Blue curves) and numerical
results (Red curves) for the case of $\omega_{\mathrm{p}}=2\pi\times4.363%
\operatorname{GHz}%
$, $a=0.017$, $T_{\mathrm{eff}}=75%
\operatorname{K}%
$, and $\omega_{\mathrm{m}}=2\pi\times1.2%
\operatorname{MHz}%
$ for panel $\left(  \mathrm{A}\right)  $ and $\omega_{\mathrm{m}}=2\pi\times2%
\operatorname{MHz}%
$ for panel $\left(  \mathrm{B}\right)  $. Panel $\left(  \mathrm{A}\right)
$, shows the reflected power, obtained for three gradually increased pump
power values,\ and corresponding to sub-harmonics generation (SHG) of the
second, third, and forth orders. SHG of the third order, for example, are
generated by a quasi-periodic response of the system (subplots $\left(
\mathrm{a}{\small 1},\mathrm{a}{\small 3}\right)  $). Each quasi-period lasts
three modulation cycles, where only during the first two a spike occurs,
namely a spike is absent once every three modulation cycles. This behavior
originates from the mismatch between the modulation period and the recovery
time of a spike, which induces a phase difference, that is monotonically
accumulated, between the two. Once every $n=3$ modulation cycles, in this
case, the system fails to achieve critical conditions near the time where the
peak in the modulation occurs, and therefore a spike is not triggered. Similar
behavior is also shown in subplots $\left(  \mathrm{a}{\small 1}%
,\mathrm{a}{\small 3}\right)  $ and $\left(  \mathrm{c}{\small 1}%
,\mathrm{c}{\small 3}\right)  $, where the quasi-period lasts two and four
modulation cycles respectively.

Another mechanism for SHG is observed when the modulation frequency is
increased. Fig. \ref{SubHarmLocking}, Panel $\left(  \mathrm{B}\right)  $,
shows measurement results for $\omega_{\mathrm{m}}=2\pi\times2%
\operatorname{MHz}%
$, which demonstrate SHG of order $n=3$. Unlike the previous case, this SHG is
characterized by a single spike that occurs once every three modulation
cycles.%
\begin{figure}
[ptb]
\begin{center}
\includegraphics[
height=2.5313in,
width=3.3723in
]%
{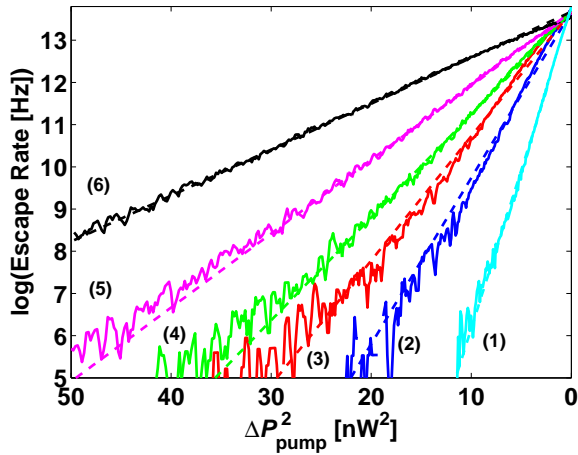}%
\caption{Escape rate of metastable states for several levels of
$T_{\mathrm{eff}}$, summarized in table \ref{EscapeRateParameters}. The graphs
are plotted in pairs, where the solid curves show the experimental data and
the dashed curves show the corresponding theoretical fit.}%
\label{EscapeRate}%
\end{center}
\end{figure}

We further study our system by measuring the fluctuation-induced escape rate
$\Gamma$ of the metastable state in the MS(S) zone. In Ref. \cite{baleegh06d}
we have found theoretically that%
\begin{equation}
\Gamma=\Gamma_{0}\exp(-\frac{\gamma_{1}\Delta P_{\mathrm{pump}}^{2}}%
{\gamma^{2}k_{\mathrm{b}}T_{\mathrm{eff}}P_{\mathrm{pump}}}),
\label{escapeRateScale}%
\end{equation}
where $\Gamma_{0}=\sqrt{H\gamma/C}/2\pi$, and the power difference is given by
$\Delta P_{\mathrm{pump}}$ $\equiv P_{\mathrm{c}}-P_{\mathrm{pump}}$. Note
that the unusual scaling law in the present case $\log\left(  \Gamma
/\Gamma_{0}\right)  \propto\Delta P_{\mathrm{pump}}^{2}$, which differs from
the commonly obtained scaling low of $\log\left(  \Gamma/\Gamma_{0}\right)
\propto\Delta P_{\mathrm{pump}}^{3/2}$ \cite{escRate_Dykman04,escRate_Biar05},
is a signature of the piecewise linear dynamics of our system.

The escape rate was experimentally measured for several levels of
$T_{\mathrm{eff}}$, which are given in the first row of table
\ref{EscapeRateParameters}. The noise was generated by an external white noise
source, and combined with the amplitude modulated pump tone. The modulation
frequency was set to $500%
\operatorname{Hz}%
$, which is more than three orders of magnitude lower than the relaxation rate
of the system, and therefore to a good approximation the system follows this
modulation adiabatically \cite{escRate_Dykman04}.

The results are shown in Fig. \ref{EscapeRate}, which plots the escape rate in
logarithmic scale as a function of $\Delta P_{\mathrm{pump}}^{2}$. Six pairs
of solid and dashed curves are shown, corresponding to the six different
levels of injected noise intensities. The solid curves were extracted from
time domain measurements of the reflected power. The dashed curves were
obtained by numerically fitting the experimental data to Eq.
(\ref{escapeRateScale}) and show good quantitative agreement between the
experimental results and Eq. (\ref{escapeRateScale}). The fitting parameters
included the pre-factor $\Gamma_{0}=0.86%
\operatorname{MHz}%
$ that was determined by a separate fitting process, and $P_{\mathrm{c}}$ (see
table \ref{EscapeRateParameters}) that slightly decreases with the thermal
noise. This behavior can be explained by local heating of the microbridge,
induced by the noise that is injected into the resonator through additional
resonance modes. Note that $T_{\mathrm{eff}}$ was extracted from a direct
measurement of the injected noise intensity (see table
\ref{EscapeRateParameters}). Note also that the system recovery time at the
threshold imposes a limit on the measured escape rate. Thus the escape rate
close to the threshold might be higher than measured.\begin{table}[ptb]
\caption{Escape rate parameters}%
\label{EscapeRateParameters}
\renewcommand{\arraystretch}{1.5}
\par
\begin{center}
$\ \
\begin{array}
[c]{c|c|c|c|c|c|c|c}
& 1 & 2 & 3 & 4 & 5 & 6 & \text{Note}\\\hline
T_{\mathrm{eff}}[10^{5}\operatorname{K}] & 0.52 & 1 & 1.36 & 1.64 & 2.3 &
3.76 & \text{Measured}\\\hline
P_{\mathrm{c}}^{\mathrm{fit}}[\operatorname{nW}] & 125.1 & 121.6 & 120.5 &
120 & 119.2 & 117.6 & \text{Fitted}%
\end{array}
$
\end{center}
\end{table}

In summary, a novel mechanism of SR with a single metastable state has been
demonstrated. Near BT the system exhibits rich dynamical effects including
bifurcation amplification and SHG. In spite of its simplicity, our theoretical
model successfully accounts for most of the experimental results.

\begin{acknowledgments}
We thank Steve Shaw and Mark Dykman for valuable discussions and helpful
comments. This work was supported by the German Israel Foundation under grant
1-2038.1114.07, the Israel Science Foundation under grant 1380021, the Deborah
Foundation, the Poznanski Foundation, Russel Berrie nanotechnology institute,
and MAFAT.
\end{acknowledgments}

\bibliographystyle{apsrev}
\bibliography{Bibilography}

\end{document}